\documentclass[superscriptaddress, aps, pre, twocolumn, floatfix ]{revtex4-2}

\usepackage{soul}
\usepackage{graphicx} 
\graphicspath{{./images/}} 
\usepackage{mathtools}
\usepackage[inline]{enumitem}

\usepackage{tikz}
\usepackage{pgfplots}
\pgfplotsset{width=10cm,compat=1.9}
\usetikzlibrary{arrows.meta}
\usetikzlibrary{patterns, shadings}

\usepgfplotslibrary{external}
\usepackage[most]{tcolorbox}
\usepackage{amsmath,amssymb}
\usepackage{booktabs}   

\usepackage[hidelinks]{hyperref}
\hypersetup{
    pdfstartview=FitH,
    colorlinks=true,
    linkcolor=blue,
    citecolor=blue,
    urlcolor=cyan
    }

\newcommand{\avg}[1]{\left< #1 \right>} 

\newcommand{\eq}[1]{Eq.~(\ref{eq:#1})} 
\newcommand{\fig}[1]{Fig.~\ref{fig:#1}} 
\newcommand{\sectn}[1]{Sec.~(\ref{sec:#1})} 

\usepackage{dcolumn}
\usepackage{bm}

\usepackage{mathtools}
\usepackage{xcolor}

\def\bea{\begin{eqnarray}}
\def\eea{\end{eqnarray}}
\def\ba{\begin{array}}
\def\ea{\end{array}}

\def\la{\langle}
\def\ra{\rangle}

\def\o{\omega_d}
\def\lmd{\lambda}

\usepackage{color}
\definecolor{dgreen}{rgb}{0,0.7,0}

\begin{document}


\title{
Observation of multiple attractors and diffusive transport in a periodically driven Klein-Gordon chain
}

\author{Umesh Kumar}
\email{umesh.kumar@icts.res.in}
\address{
International Centre for Theoretical Sciences, 
Tata Institute of Fundamental Research, 
Bengaluru 560089, India
}

\author{Seemant Mishra}
\email{seemant.mishra@uni-osnabrueck.de}
\address{
Universit\"at Osnabr\"uck, Faculty of Mathematics, Informatics and Physics,
Institute of Physics, Barbarastra{\ss}e 7, D-49076 Osnabr\"uck, Germany
}

\author{Anupam Kundu}
\email{anupam.kundu@icts.res.in}
\address{
International Centre for Theoretical Sciences, 
Tata Institute of Fundamental Research, 
Bengaluru 560089, India
}

\author{Abhishek Dhar}
\email{abhishek.dhar@icts.res.in}
\address{
International Centre for Theoretical Sciences, 
Tata Institute of Fundamental Research, 
Bengaluru 560089, India
}

\date{\today}

\begin{abstract}
We consider  a Klein-Gordon chain that is periodically driven at one end and has dissipation at one or both boundaries.  An interesting numerical observation in a recent study~\cite{prem2023} was that for driving frequency in the phonon band, there is a range of values of the driving amplitude $F_d \in (F_1, F_2)$ over which the energy current remains constant. In this range, the system exhibits a traveling wave solution  termed as a ``resonant nonlinear wave" (RNW). It was noted that  the RNW mode occurs over a range $(F_1, F_2)$ and shrinks with increasing system size, $N$. Remarkably, we find that the RNW mode is in fact a stable solution even for $F_d >F_2$, 
and that  in this regime there exists two  attractors both with finite basins of attraction.  
We improve the perturbative treatment of Ref.~\cite{prem2023} for the RNW mode by including the contributions of  third harmonics. We also consider the effect of thermal noise at the boundaries and find that the RNW mode is stable for small temperatures. Corresponding to the two attractors for large $F_d$ at zero temperature, the system can now be in two nonequilibrium steady states.
Finally, we present  results for a different driving protocol studied in Ref.~\cite{komorowski2023} where $F_d$ is taken to scale with system size as $N^{-1/2}$ and  dissipation is only at the non-driven end. We find that the steady state can be characterized by Fourier's law as in Ref.~\cite{komorowski2023} for a stochastic model. We point out interesting differences that occur because of our dynamics being nonlinear and Hamiltonian. Our results suggest the  intriguing possibility of observing the high current carrying RNW phase in experiments by careful preparation of initial conditions.
\end{abstract}

\maketitle



\section{Introduction}
\label{sec:intro}
Isolated many-particle systems, described by Hamiltonians with generic nonlinear interactions, typically have few conservation laws, display chaotic dynamics, and it is expected that at long times they should show ergodic behavior where statistically the system is well described by equilibrium Gibbs ensembles.
In the presence of boundary driving and dissipation, these systems evolve to nonequilibrium steady states (NESS). One of the most widely studied setup is
when the two ends of a 
chain of $N$ oscillators
are 
coupled to heat baths at different temperatures. For the case where all interactions are harmonic~\cite{1967RLL}, one has ballistic transport where the NESS energy current $J$ is independent of system size $N$ (for sufficiently large $N$). For anharmonic systems with external pinning potentials, transport is in accordance with Fourier's law~\cite{2019dhar} with $J\sim N^{-1}$ while for anharmonic chains with momentum conservation one gets anomalous transport~\cite{2003LLP,dhar2008,2020benenti} with $J\sim N^{\alpha-1}$, where $0<\alpha<1$. A second setup of interest is one where the system is driven through a boundary periodic force instead of thermal noise. This has been well studied in the context of supra-transmission~\cite{geniet2002, geniet2003,khomeriki2004}, which refers to transmission of energy at driving frequencies outside the phonon bandwidth, observed beyond some critical driving amplitude. Other interesting results have been obtained in the context of thermal ratcheting~\cite{li2008} and non-reciprocal transmission~\cite{narayan2004,lepri2011}. Another relevant class of studies that has attracted a lot of attention are Floquet systems~\cite{bukov2015,higashikawa2018floquet} where typically one considers bulk periodic driving --- the boundary driving that we consider here constitutes a new interesting class. 

Most recently, energy transmission in a periodically driven Klein-Gordon (KG) chain was studied~\cite{prem2023} in the context of experiments on photon transmission in arrays of quantum oscillators~\cite{fitzpatrick2017,fedorov2021}. We will here refer to the work in Ref.~\cite{prem2023} as the Prem-Bulchandani-Sondhi (PBS) setup. This work considered a KG chain of  $N$ particles with dissipation at both ends and   one end of the chain  being driven by a sinusoidal force $F(t)=F_d\cos(\omega_d t)$. As a result of the driving, the system reaches a non-equilibrium steady state (NESS) which is characterized by an average energy current $J$ flowing from the driven end to the non-driven end. 
With the driving frequency ($\omega_d$) kept fixed at some value in the band of the  underlying harmonic chain (neglecting the anharmonic terms), the authors in \cite{prem2023} noted interesting transitions as one varied the driving force strength $F_d$, for a long but finite chain. On increasing $F_d$ from $0$, the system first transits from a chaotic to a periodic state, which is stable in the regime $F_{1}<F_d<F_{2}$, after which the system again moves to a chaotic state.
In the range $F_{1}<F_d<F_{2}$, the mean energy current was found to be independent of $F_d$ and $N$ (for not very large $N$), and depended on the driving frequency $\omega_d$ in a non-trivial manner.  The periodic state was identified as a Resonant Nonlinear Wave (RNW) where the bulk particles ($1<<\ell<<N$) have the form, $q_\ell= 2 r \cos (\omega_d t-k\ell)$, characterized by a constant phase difference $k$ between successive particles and with the amplitude $r$ given explicitly by 
\begin{equation}
\label{eq:bulk_amp}
r = \sqrt{\frac{\omega_d^2-1 - 2(1-\cos{k})}{3}}.
\end{equation}
In simulations, the  RNW mode was not seen for  $F_d>F_2$ and it was noted that the system reached a chaotic NESS. This NESS exhibits  a Fourier-like scaling,  $J \sim N^{-1}$, for the current, though surprisingly, the chain had a large segment where there was no local equilibrium.


\begin{figure*}
\includegraphics[width=\linewidth]{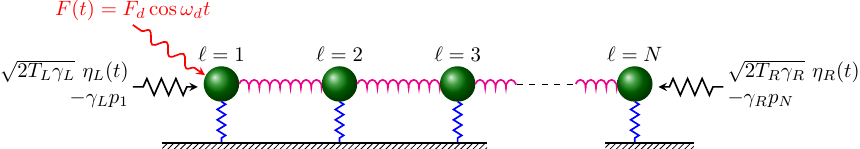}
\caption{(color online)
Schematic diagram of  the driven Klein-Gordon chain. Springs represent interaction as well as external potential present in the Hamiltonian given in Eq.~\eqref{H_kg}. The system is attached to two heat baths at the two ends and is periodically driven at the left end.}
\label{fig:chain}
\end{figure*}


In another interesting  recent work~\cite{komorowski2023}, a harmonic chain with a stochastic energy conserving dynamics was periodically driven at one end, and had  dissipation and thermal noise applied only at the other end [we refer to this as the Komorowski-Lebowitz-Olla (KLO) setup]. The stochastic part of the dynamics conserved energy, but not momentum. It was shown that in this case, choosing $F_d\sim N^{a}$ and $\omega_d \sim N^b$ with $b-a=1/2$, led to a unique periodic stationary state where the average current is in accordance with Fourier's law (for large $N$). The temperature profile could be obtained from the  diffusion equation with Neumann boundary condition at the driven end and Dirichlet at the other. The problem in the absence of the stochastic bulk noise has also been studied~\cite{pedro2023klo}.

In this work, we investigate further the  RNW mode, in particular the question of  its stability and the effect of thermal noise that inevitably accompanies dissipation.   Secondly, we investigate the KLO setup where the periodic boundary driving is of the form $F(t)=\frac{A}{\sqrt{N}}\cos(\omega_d t)$ with $A$ being a constant. This corresponds to  $a=-1/2$ and $b=0$ in Ref.~\cite{komorowski2023}.  We emphasize that, unlike the  study of \cite{komorowski2023}, the bulk dynamics considered by us is purely Hamiltonian. An interesting question that arises from the KLO paper is whether we can effectively describe the periodically driven system as one satisfying the heat diffusion equation in the bulk but with a Neumann boundary condition at the driven end. In our studies, in order to verify if the temperature profiles satisfied the heat diffusion equation (with a temperature dependent conductivity) we compared the temperature profile with that obtained from simulations of a chain driven purely thermally at both ends (with no periodic driving). The boundary temperatures of the thermally driven chain were chosen to correspond to temperatures at  points, away from the boundary jumps of the periodically driven chain.

We summarize the main results of our work for the PBS and KLO setups:
\begin{itemize}

\item {\bf PBS: Stability of RNW mode in PBS setup} - We verify that for generic initial conditions, the system evolves at long times to the RNW mode for $F_1<F_d<F_2$, as was observed in \cite{prem2023}. However, we demonstrate,  that the RNW is in fact stable even in the regime $F_d>F_2$. It can be reached by starting from  by starting from initial conditions close to an RNW.   We also show that the RNW mode (for $F_d>F_2$) has a finite basin of attraction, which we numerically estimate by finding the minimum perturbation required to push it to the chaotic state.  Surprisingly, our numerical results indicate that the  size  of the basin of attraction converges to a finite value (see Fig.~\ref{fig:f1f2-rnw-crit-perturb}) for increasing values of the driving force $F_d$.  Thus, we conclude that for all $F_d>F_2$, the dynamical system has two attractors, one chaotic and the other periodic. Our results suggest that there is a unique periodic attractor in the range $F_1<F_d<F_2$ and a unique chaotic attractor in the range  $F_d<F_1$. 

\item {\bf PBS: Third harmonic contributions to RNW} - It was shown in \cite{prem2023} that RNW is well described in the bulk by the form $q_\ell= 2 {\rm Re} [r  e^{\iota(\omega_d t-k\ell)}] $ with $r$ given in \eq{bulk_amp} and $k$ a numerically determined constant.   Near the edges there are boundary layers in which the amplitude and the phase difference are site-dependent. In this work, we improve the proposed form of the RNW by including  the contribution of the $3^\text{rd}$ harmonic. More precisely, we assume the solution $q_\ell = {\rm Re}[a_\ell e^{\iota\omega_d t} + b_\ell e^{\iota3\omega_d t}]$. Inserting this in the original dynamical equation and neglecting harmonics at $5 \omega_d$, we get a set of algebraic equations for $\{a_\ell,b_\ell\}$ which we solve numerically. We find that, in the bulk, the complex amplitudes can be written as $a_\ell = r_{1} e^{-\iota k_1 \ell}$ and $b_\ell = r_{3} e^{-\iota 3k_1 \ell}$ with  $r_{1}$ and $r_{3}$ are given by Eq.~\eqref{eq:r1r3-3wd}. The amplitude profile at the boundary approaches  the bulk value exponentially  with a rate that is independent of  system size (see  Fig.~\ref{fig:r-vs-l-without-3wd-N=100,200}).

\item {\bf PBS: Effect of Noise} - We  study the effects of boundary thermal noise on the stability of the RNW. We find that the sharp transitions in the current with changing $F_d$ persists for small temperatures but disappears at high temperatures. In this case, the system is expected to go into a time-periodic steady state. Our results indicate that the signatures of multiple attractors seen for the zero-noise case continue to persist for the low noise case and there could be non-unique steady states (see Fig.~\ref{fig:rnw-at-finite-temp}). At higher temperatures the transitions go away, and we ask whether we obtain diffusive transport, following Fourier's law. 
{For this, we computed the  kinetic temperature profiles and the mean current for the case where the boundary thermal temperature is large. We find that a big temperature jump appears between the first and second sites, while the bulk profile is smooth and appears to satisfy Fourier's law (from comparisons with a thermally driven chain). The current shows a $N^{-1}$ scaling with system size.}

\item {\bf Other boundary conditions in the presence of noise:}
Here we considered the case where the left end  has periodic driving {\emph{but no dissipation}}, while the right end is driven by a thermal bath. We discuss two cases:\\ \\
{\emph{(i) $F_d \sim \frac{A}{\sqrt{N}}$ (KLO setup)}:} This is the setup discussed in Ref.~\cite{komorowski2023} but with  bulk dynamics being Hamiltonian, in contrast to the stochastic dynamics of KLO.  In this case, the temperature profile seems to converge to a limiting form while the current still decays as $N^{-1}$. The temperature profile has  a jump between first and second sites, which decreases with increasing system size.  We find that for the largest system size, the  temperature profiles as well as the current of the thermally driven chain match with the periodically driven case, thus suggesting validity of Fourier's law in this system.  

{\emph{(ii) $F_d$ independent of $N$:}} In this case, we find that the temperature profile does not attain a limiting form for increasing $N$. Instead, the temperature of the left end diverges as $N^{1/2}$. The current still has a $N^{-1}$ dependence, and the temperature profile seems to be in accordance to the heat diffusion equation (from comparisons with a thermally driven chain).

\end{itemize}

The rest of the article is organized as follows. In \sectn{model} we define the precise model, describe the observables that we study and the numerical methods used. In Sec.~\eqref{sec:T=0} we discuss the noiseless case where we provide numerical evidence for the existence of the RNW mode beyond the second transition at $F_2$, study its stability, and discuss the corrections to the form of the RNW mode arising from third harmonic contributions. Finally, in \sectn{T!=0} we study the behavior of the system in the presence of thermal noise at the ends of the chain, for the PBS setup and also the KLO setup. We end with our conclusions in \sectn{conclusion}.

\begin{figure}
\includegraphics[width=\linewidth]{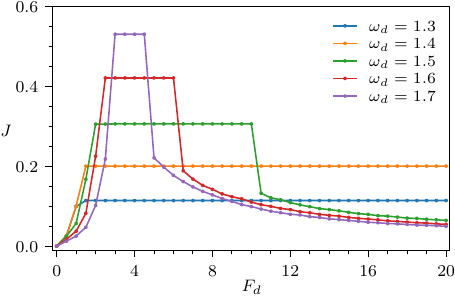}
\caption{{\bf PBS setup ($T_L=T_R=0$):} Variation of steady state current $J$ with driving amplitude $F_d$ for system size $N=500$ at different driving frequencies $\omega_d$ inside the harmonic band $(1,\sqrt{5})$.}
\label{fig:JvsF-and-wd-at-T0}
\end{figure}

\section{Model and Observables}
\label{sec:model}

We consider a chain of $N$ particles, each with mass $m$, where the position and momentum of the $\ell^{th}$ particle are respectively denoted by $q_\ell$ and $p_\ell$. The Hamiltonian of the chain is:
\begin{equation}
H = \sum_{\ell=1}^N \frac{p_\ell^2}{2m} 
+ \frac{m \omega_0^2}{2} q_\ell^2 + \frac{\nu}{4} q_\ell^4 
+ \sum_{\ell=1}^{N-1} \frac{\lambda}{2} (q_{\ell+1} - q_\ell)^2,
\label{H_kg}
\end{equation}
In addition to  the Hamiltonian dynamics, the system is  connected to the baths at the two ends  which we model through Langevin equations having dissipation and noise terms, corresponding to  baths at temperatures $T_L$ and $T_R$.   Finally, the chain is also driven by a periodic force at the left end with amplitude $F_d$ and frequency $\omega_d$. A schematic figure of the model is shown in \fig{chain}. The resulting equations of motion for the system  are thus:
\begin{align}
&m\ddot{q}_\ell =
-m\omega_o^2 q_\ell -\nu q_\ell^3 
+ \lambda (q_{\ell+1}+q_{\ell-1}-2q_\ell) 
\nonumber \\ &+ \delta_{\ell,1} \Biggl(-\gamma_L\dot{q}_\ell 
+  \sqrt{2k_B T_L\gamma_L}~\eta_L(t) 
+ F_d\cos(\omega_d t)\Biggr) \nonumber \\ &+ \delta_{\ell,N} \Biggl(-\gamma_R\dot{q}_j
+  \sqrt{2k_B T_R \gamma_R}~\eta_R(t)\Biggr),~~~\ell=1,\dots,N,
\end{align}
where we consider free boundary conditions $q_0=q_1,~q_{N+1}=q_N$ and the thermal noises, $\eta_{L/R}(t)$, have zero mean and variances  $\avg{\eta_a(t) \eta_b(t')}= \delta_{a,b}~\delta(t-t')$ for $a,b\in \{L,R\}$.
 Let us use dimensionless variables by rescaling time and position as
\begin{equation}
\omega_o t\to t\, \quad \text{and} \quad  \sqrt{\frac{\lmd }{m \omega_0^2}}q_\ell \to q_\ell,
\end{equation}
which results in the following  equations of motion:
\begin{align}
&\ddot{q}_\ell = -q_\ell -q_\ell^3 + \lambda (q_{\ell+1}+q_{\ell-1}-2q_\ell) \nonumber \\
&+ \delta_{\ell,1} \Biggl(
- \gamma_L~\dot{q}_1 
+ \sqrt{2T_L \gamma_L}~\eta_L(t) 
+ F_d\cos(\omega_d t)\Biggr) \label{eq:eom_s}\\
&+ \delta_{\ell,N} \Biggl(-
\gamma_R~\dot{q}_N
+ \sqrt{2T_R\gamma_R}~\eta_R(t)
\Biggr),~~~\ell=1,\dots,N, \nonumber
\end{align}
where the rescaled system-bath parameters are transformed as:
\begin{align}
 \frac{\lambda}{m \omega_o^2} &\to \lambda
\quad
\frac{\gamma_{L,R}}{m \omega_o} \to \gamma_{L,R},
~~
 \frac{\omega_d}{\omega_o} \to \omega_d \nonumber\\
 \frac{\nu^{1/2}}{(m\omega_o^2)^{3/2}} F & \to F_d,~
\frac{k_B\nu}{m^2\omega_o^4}T_{L,R} \to T_{L,R},~
 \frac{\eta_{L,R}}{{\omega_o}^{1/2}}  \to \eta_{L,R}. \nonumber
\end{align}
We will only typically look at the behavior of the system as the parameters $F_d$, $\o$, $N$, and $T_{L,R}$ are varied while $\lambda$ and $\gamma_{L,R}$ are set to the value one (for KLO setup $\gamma_L=0$). Note that the spectrum of the harmonic part is given by $\Omega=\sqrt{1+2 \lambda (1-\cos q)}$ and so lies in the range $(1,\sqrt{1+4\lambda})$. In all our computations we set $\lambda=1$ which correspond to the harmonic band $(1,\sqrt{5})$.

To understand the behavior of the system in its steady state, we  look at the steady state current and local kinetic temperature, which are given by 
\begin{subequations}
\begin{align}
J_\ell &= \avg {\lambda (q_{\ell-1}-q_\ell) p_\ell} , \label{eq:curr}\\
T_\ell &= \langle p_\ell^2 \rangle,
\label{eq:temp}
\end{align}
\end{subequations}
where $\la \cdot \ra$ denotes a time average in the steady state. More precisely, we compute the following time average of any observable, $A$, :
\begin{align}
\avg {A} = \frac{1}{\tau} \int_{\tau_0}^{\tau_0+\tau} dt~ A(t), 
\end{align}
for very large $\tau$ and $\tau_0$. We study the behavior of the current and temperature profile as a function of the system size $N$ and driving parameters $F_d, \o$ and $T_{L,R}$. In the following sections, we consider  different periodic driving protocols. We also consider a thermal driving protocol where we set $F_d=0$ and impose different temperatures $T_L \neq T_R$ at the two ends.

{\bf Numerical methods}: For our noiseless  simulations, we dynamically evolve  \eq{eom_s} using the  velocity-Verlet algorithm \cite{allen-tildesley} with time step $dt=0.01$, and measure the relevant observables in the steady state. To push the system into the steady state we first evolved the system for $R_{trans}=\tau_0/dt$  number of transient time steps, and then we  computed steady state averages from data over the next $R_{avg}=\tau/dt$ steps. We verified that system had reached  the steady state by ensuring that a flat current profile is attained. For most of our investigations, we used $R_{trans}=10^8$ and $R_{avg}=2\times 10^8$ for chain of size $N<1600$, while for system size $N\ge 1600$ we used larger number of averaging steps, namely  $R_{avg}=10^9$ for $N=1600$ and $R_{avg}=10^{10}$ for $N=3200$. The simulations with noise were performed using the stochastic velocity-Verlet~\cite{allen-tildesley}, again with $dt=0.01$. The numerical solution of \eq{ab-b-eqn} were found using SciPy's root finding routine\cite{scipy}. For finding the solution we used the profiles of $r$ and $\varphi$ found through simulation.

{\bf Equivalent thermal drive}: The simulations of a chain driven purely thermally was done by setting $F_d=0$ and $T_L \neq T_R$.   The boundary temperature, $T_R$ was set to the value $T$ at the site $\ell=N$.   The temperature, $T_L$,  at the left end of the thermally driven chain was chosen to correspond to temperatures at  points, away from the  temperature jump at the left boundary of  the periodically driven chain. 

\section{PBS: Results for the noiseless case}
\label{sec:T=0}

Here we consider the  case without noise ($T_L=T_R=0$) studied by PBS~\cite{prem2023}.   Taking $\lambda=1$, 
$\gamma_L=\gamma_R=1$ and $T=0$, the equation of motion  becomes:
\begin{align}
\label{eq:eom_T0}
\ddot{q_\ell} 
&= -q_\ell - q_\ell^3 + (q_{\ell+1}+q_{\ell-1}-2 q_\ell) \nonumber\\
&\qquad\qquad\quad
+\delta_{\ell, 1}(-\dot{q_\ell} 
+ F_d\cos\omega_d t)
-\delta_{\ell, N}(\dot{q_\ell}),
\end{align}
for $\ell =1,2,\ldots,N$ with $q_0=q_{N+1}=0$. We first discuss the case where we start from random initial conditions (IC) and evolve the system for a long time to reach the steady state. In \fig{JvsF-and-wd-at-T0}, we show the steady state current as a function of the driving force for different values of the driving frequency for a chain of size $N=500$. This  reproduces one of the most interesting results of \cite{prem2023}, namely the observation of the current plateau over a range of force values $F_1< F_d< F_2$. We also observe from the figure that $F_1$ and $F_2$ are functions of frequency. We find that the plateau is observed only when $\omega_d$ belongs to the harmonic chain phonon band [$\omega_d \in (1,\sqrt{5})$]. 

\subsection{Observation of RNW beyond the transition point $F_2$}
\begin{figure}[h]
\includegraphics[width=\linewidth]{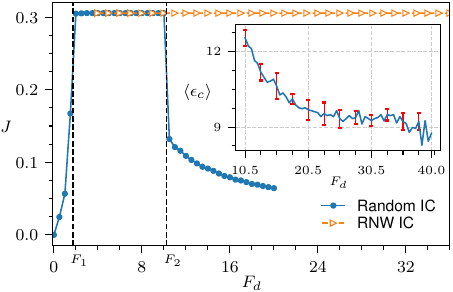}
\caption{(color online) {\bf PBS setup ($T_L=T_R=0$):}
Variation of steady state current $J$ with driving amplitude $F_d$ at driving frequency $\omega_d=1.5$ and system size $N=500$. The system is initialized either with Random IC (blue dots) or with RNW IC (orange triangle).
With random IC we observe transitions, to and from the RNW mode, at values of $F_d$ indicated by the vertical black dashed lines. For RNW IC, we observe that the RNW phase continues beyond $F_2$. As shown in the inset,  the RNW mode beyond $F_2$ is  stable against adding perturbations. The critical perturbation strength $\langle \epsilon_c \rangle$ required to destabilize the RNW mode for different $F_d$ is plotted (inset) and we observe that $\langle \epsilon_c \rangle$ seems to saturate at large $F_d$. This seems to suggest that the size of basin of attraction of the RNW mode remains finite for arbitrarily large $F_d$.}
\label{fig:f1f2-rnw-crit-perturb}
\end{figure}

We now report our first main result, which is a numerical demonstration that in fact the RNW mode continues to be a stable solution, even when the driving force is larger than $F_2$. To observe the RNW mode beyond, $F_2$ it is necessary that we \emph{not} start from a random IC. Instead, we increase the force in small steps ($F_d \to F_d+\Delta F$). At the new step (with $F_d+\Delta F$) we use as initial conditions the set of position and momenta values from the last time of the previous simulations (at $F_d$), i.e, we always start from initial conditions which are close to the RNW. We then find that the system current continues to be on the plateau even for forces as large as $F_d\approx 40$ which is much beyond the transition value $F_2\approx 10$ observed when we start from random initial conditions. This is shown in \fig{f1f2-rnw-crit-perturb}. This suggests that for $F_d>F_2$, the long time dynamics of the system has two attractors, one of which is chaotic (obtained by starting from random IC) and the other a periodic state corresponding to the RNW. We now estimate  the size of the basin of attraction of the RNW. We probe this by adding random perturbations to the initial state of the RNW (specified by $\{q_j,p_j\}$ of the form $q_j' = q_j + \epsilon s^q_j,~p_j' = p_j + \epsilon s^p_j$, for $j=1,2,\ldots,N$, where ${\bf s}=\{s^q_j,s^p_j\}$  is a  random unit vector on a $2N$-dimensional unit sphere.   Given a value of $\epsilon$ and any realization, ${\bf s}$ we evolve the system to see if it goes to the chaotic state. Since the basin of attraction can have a highly irregular shape, the value $\epsilon_c$ at which the system becomes unstable, depends on the direction ${\bf s}$. Hence, for each $F_d$, we compute $\langle \epsilon_c \rangle$ by averaging over $10$ random directions. In the inset of Fig.~\ref{fig:f1f2-rnw-crit-perturb} we show a plot of  $\langle \epsilon_c \rangle$ as a function of the driving amplitude $F_d$ for $N=500$ and $\o =1.5$. We observe a  decrease in $\langle \epsilon_c \rangle$ as $F_d$ is increased, indicating that the size of the basin of attraction decreases. However, $\langle \epsilon_c \rangle$ seems to saturate to  a constant value, suggesting that the size of the basin of attraction remains finite for arbitrarily large values of~$F_d$. 
\begin{figure}[h]
\includegraphics[width=\linewidth]{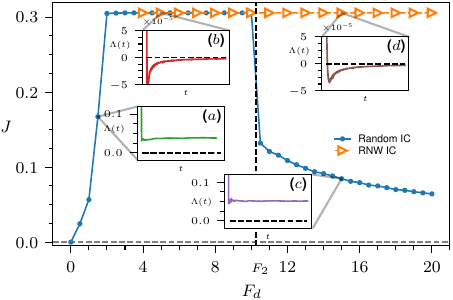}
\caption{(color online) {\bf PBS setup ($T_L=T_R=0$):} The insets  display the  Lyapunov exponents as functions of time   in different  regimes  indicated in the main plot (same as in \fig{f1f2-rnw-crit-perturb}, for system size $N=500$ and driving frequency $\omega_d=1.5$. As expected, the Lyapunov exponents saturate to zero for RNW modes and to finite positive values in the  chaotic regimes.
}
\label{fig:lyapunov_exp}
\end{figure}
In Fig.~\ref{fig:lyapunov_exp} we plot the largest Lyapunov exponent ($\Lambda$) in different force regimes. For $F_d>F_2$ we see that $\Lambda$ vanishes for initial conditions corresponding to the RNW, while for generic initial conditions we get a non-zero $\Lambda$ implying a chaotic attractor. As mentioned earlier for $F_1<F_d<F_2$, the system starting from a generic IC reaches the RNW state after long time. Consequently, the $\Lambda$ in this regime approaches to  zero at large time as expected. On the other hand, the Lyapunov exponent for $F_d<F_1$ always saturates to a non-zero value indicating chaotic behavior.

\subsection{PBS: Third harmonic contribution to RNW solution}

As noted in \cite{prem2023}, we can gain some insight about the RNW by looking for approximate analytic solutions of the nonlinear equations of motion. In particular, plugging the ansatz,  
 \begin{align}
    q_\ell = {\rm Re} (a_\ell e^{\iota \o t}),
\label{eq:ans1}
\end{align}
into \eq{eom_T0}, and upon equating  terms proportional to $e^{i \omega_d t}$, while  neglecting higher frequency terms of the form $e^{\pm \iota 3\o t}$, we get 
\begin{align}
\label{eq:as1}
(1-\omega_d^2)a_\ell 
&+ 3|a_\ell|^2a_\ell +(2a_\ell -a_{\ell+1}-a_{\ell-1})  \\
&+ \delta_{\ell, 1} (\iota\omega_d a_\ell -F_d/2)
+ \delta_{\ell, N} (\iota\omega_d a_\ell) = 0, \nonumber
\end{align}
for $\ell=1,2\ldots N$, where we have taken the boundary conditions $a_0=a_1$ and $a_{N+1}=a_N$. This is a non-linear set of equations which can be  solved numerically to obtain solutions in the form  $a_\ell = r_\ell e^{\iota \varphi_\ell}$ with real $r_\ell$ and $\varphi_\ell$. It was observed in PBS that for sites $\ell$ in the bulk, the amplitudes were constant and the phase difference was constant, i.e, $r_\ell=r$ and  $\varphi_\ell-\varphi_{\ell+1}= k$ where $r$ and $k$ are $\ell$-independent. Plugging  the form $a_\ell=r e^{-\iota k \ell}$ in \eq{ans1} for the bulk points, one obtains 
Eq.~\eqref{eq:bulk_amp}. 

\begin{figure}[h]
\includegraphics[width=\linewidth]{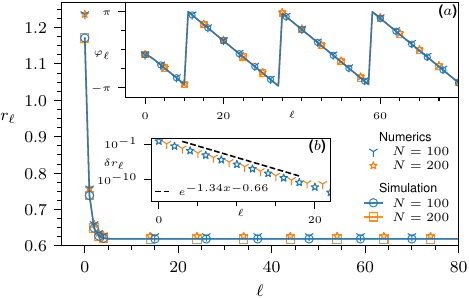} 
\caption{(color online) {\bf PBS setup ($T_L=T_R=0$):} 
Plots of the profiles of amplitude $r_\ell$ and phase $\phi_\ell$ (inset a) calculated from the  numerical solution of \eq{as1} and extracted from simulation of \eq{eom_T0}, for system sizes $N=100,200$, with $F=10.0$ and $\omega_d=1.5$. Note the slight difference in $r_l$ between  numerical solution ( dashed orange line) and simulation (blue solid line). The inset (b) shows the exponential decay of the deviation $\delta r_\ell =r_\ell -r_{\rm bulk}$. }
\label{fig:r-vs-l-without-3wd-N=100,200}
\end{figure}

A non-zero value of $k$ implies a traveling wave solution and is  important to get a finite value of current. From Eq.~\eqref{eq:curr} one finds the time-averaged current to be given by~\cite{prem2023}:
\bea \label{eq:j_form}
J = 2 r^2 \o \sin{k}.
\eea

The bulk solution does not determine the phase difference $k$ and it could potentially depend on the parameters $F_d, \omega_d$ and $N$. A full solution of Eq.~\eqref{eq:as1} including the boundary conditions would of course also determine completely all $r_\ell$ and $\varphi_\ell$ (and hence the constant phase difference $k$ in the bulk). In \fig{r-vs-l-without-3wd-N=100,200} we compare the results for $r_\ell$ and $\varphi_\ell$ [see inset (a)], obtained  from a numerical solution of Eq.~\eqref{eq:as1},  with those obtained from direct simulations of \eq{eom_T0} and find quite good agreement but also observed some discrepancy which is most prominent at the driven end. The inset $(a)$ of the figure shows the exponential decay of the amplitude to the bulk value at the driven boundary. We also observe that the boundary profile is independent of system size. In inset $(b)$, we observe that the phase $\varphi_\ell$ changes linearly with site index $\ell$ at a rate $k$ which is independent of system size.

\begin{figure}[h]
\includegraphics[width=\linewidth]{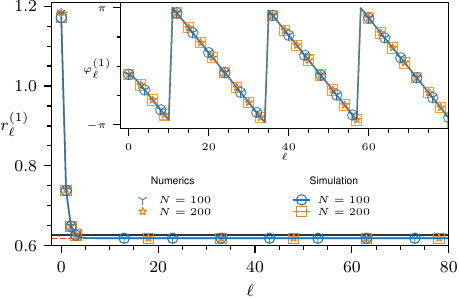}
\caption{(color online) {\bf PBS setup ($T_L=T_R=0$):} Here we plot the  amplitude and phase 
profiles computed from the solution of Eq.~\eqref{eq:ab-eqns} which includes the contributions of the $3^{\rm rd}$ harmonics. We now see a better agreement with simulation results,  as compared to the $1^{\rm st}$ harmonic numerical results discussed in Fig.~\ref{fig:r-vs-l-without-3wd-N=100,200}. To see the improvement, we also replot the numerical result (black dashed line shows) obtained from the $1^{\rm st}$ harmonic approximation.
}
\label{fig:r1-phi1-profile-with-3wd}
\end{figure}

Even though the ansatz in \eq{ans1} provides a good description of the RNW as a single frequency mode, there are some differences as pointed out above. This is due to the fact that $3\omega_d$ contribution for the first particle is much larger than the rest of the chain. To improve the resonant solution, we now incorporate the third harmonic corrections
--- specifically, we make the following ansatz,
\begin{equation}\label{eq:rnw-with-3wd}
q_\ell = {\rm Re} [a_\ell e^{\iota\omega_d t} + b_\ell e^{\iota3\omega_d t}], 
\end{equation}
which includes a  frequency response at $3\omega_d$.  Plugging this ansatz into \eq{eom_T0}, and equating coefficients of $e^{\iota\omega_d t},e^{\iota 3  \omega_d t}$ respectively to zero, we get the following sets of coupled equations for $\{a_\ell,b_\ell\}$:
\begin{subequations}
\label{eq:ab-eqns}
\begin{align}
(1-\omega_d^2)a_\ell 
&+ 3|a_\ell|^2a_\ell + 3(a_\ell^*)^2b_\ell 
+ (2a_\ell -a_{\ell+1}-a_{\ell-1})\nonumber\\
&+ \delta_{\ell, 1} (\iota\omega_d a_\ell -F_d/2)+ \delta_{\ell, N} (\iota\omega_d a_\ell) = 0
\label{eq:ab-a-eqn} \\
(1-9\omega_d^2)b_\ell 
&+ 3|b_\ell|^2b_\ell + 6|a_\ell|^2b_\ell + a_\ell^3\nonumber\\
&+ (2b_\ell -b_{\ell+1}-b_{\ell-1})\nonumber\\
&+ \delta_{\ell, 1} (\iota3\omega_d b_\ell)+ \delta_{\ell, N} (\iota3\omega_d b_\ell) = 0,
\label{eq:ab-b-eqn}
\end{align}
\end{subequations}
where we assume free boundary conditions $a_0=a_1, a_N=a_{N+1}, b_0=b_1, b_N=b_{N+1}$. We  solve these equations numerically to find $a_\ell$ and $b_\ell$. Again, we write these complex numbers in terms of their real amplitudes and phases as $a_\ell=r^{(1)}_\ell e^{\iota\varphi^{(1)}_\ell}$ and $b_\ell=r^{(3)}_\ell e^{\iota\varphi^{(3)}_\ell}$.

\begin{figure}[h]
\includegraphics[width=\linewidth]{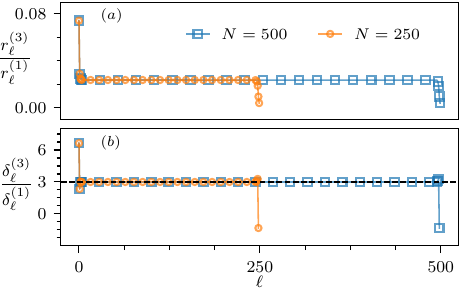}
\caption{(color online) {\bf PBS setup ($T_L=T_R=0$):} To show the relative contribution of the $3^{\rm rd}$ harmonics to $1^{\rm st}$ harmonic, we plot the ratios of the amplitudes in (a) and of the phase differences in (b) as functions of $\ell$.  We observe that both the ratios saturate to values $r_3/r_1=0.02$ and $k_3/k_1=3$.  The profiles are  calculated from  simulation of \eq{eom_T0}, for $F_d=5.5$ and $\omega_d=1.5$ for $N=250$ and $N=500$. }
\label{fig:r-and-phi-profile-forlarge-N}
\end{figure}

\begin{figure}[h]
\includegraphics[width=\linewidth]{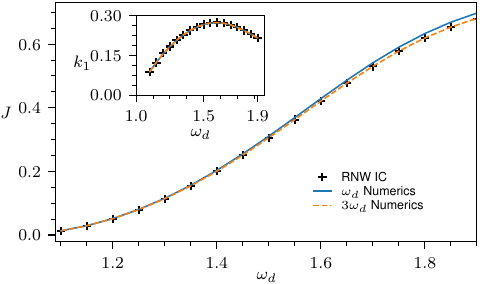}
\caption{(color online) {\bf PBS setup ($T_L=T_R=0$):} Variation of steady state current $J$ with driving frequency $\omega_d$ for system size $N=100$ and driving amplitude $F_d=10.0$.  We  compare the simulation results (plus points) with those obtained from first harmonic (solid line) and third harmonic (dashed line) numerical  computations. Note the improved agreement  of the $3^{\rm rd}$ harmonic numerical computation with simulation results. The inset shows the variation of phase difference, $k_1$, between consecutive particles  in the bulk with driving frequency $\omega_d$. In this case, there is no visible difference between the $1^{\rm st}$ (solid line) and $3^{\rm rd}$ harmonic numerical computation. They both match very well with the simulation data.
}
\label{fig:k1-vs-wd}
\end{figure}

In \fig{r1-phi1-profile-with-3wd} we show a  comparison of the values of  $r^{(1)}_\ell$ and $\varphi^{(1)}_\ell$, obtained from direct simulations with those from the numerical solution of Eqs.~\eqref{eq:ab-a-eqn} and \eqref{eq:ab-b-eqn}, and find improved agreement [compared to the first harmonic results presented in Fig.~\ref{fig:r-vs-l-without-3wd-N=100,200}]. We observe that in the bulk $r_\ell^{(1)}$ saturates to the value $r_1$. 
The linear dependence of $\varphi^{(1)}_\ell$ on $\ell$ (in the inset of Fig.~\ref{fig:r-vs-l-without-3wd-N=100,200}) suggests that the phase difference between consecutive sites $\delta^{(1)}_\ell=\varphi^{(1)}_\ell-\varphi^{(1)}_{\ell+1}$ is a constant,  {\it i.e.,} independent of $\ell$ inside the bulk, which we denote by $k_1$.   We also compute $r_\ell^{(3)}$ and $\delta_\ell^{(3)} = \varphi^{(3)}_\ell-\varphi^{(3)}_{\ell+1}$ from simulation and find that they also are $\ell$ independent inside the bulk and are denoted by $r_{3}$ and $k_{3}$ respectively. However, the constant value $r_3$ of $r_\ell^{(3)}$ in the bulk is much smaller than that of $r_1$ as can be seen from Fig.~\ref{fig:r-and-phi-profile-forlarge-N}. We also find interestingly that, $k_3 = 3 k_1$ inside the bulk (also shown in Fig.~\ref{fig:r-and-phi-profile-forlarge-N}). 
This condition immediately follows on writing the bulk equations, which then leads to the following relations between the constant amplitudes and phase differences:
\begin{align}
\begin{split}
(r_{1}^2 - r^2) + r_1 r_3  &= 0\\
(-3B^2 r_3+3r_3^3+6r_1^2r_3) + r_1^3 &= 0,
\end{split}
\label{eq:r1r3-3wd}
\end{align}
where $r$ is given by \eq{bulk_amp} and $B^2 = r^2(9+10\Delta/3)$ with $\Delta=(12-9\cos k_1+\cos 3k_1)/5r^2$. These equations can be solved to give expressions for $r_1$ and $r_3$ in terms of $\omega_d$ and $k_1$ and are equivalent to Eq.~\eqref{eq:bulk_amp}  obtained from the $1^{\rm st}$ harmonic approximation. While $r_1$ and $r_3$ can be computed analytically, the bulk phase difference $k_1$ still needs to be obtained by solving the full set of equations \eqref{eq:ab-a-eqn} and \eqref{eq:ab-a-eqn} along with the boundary conditions. Finally, incorporating all the contributions from the third harmonics, we get an improved version of Eq.~\eqref{eq:j_form} for the current given by  
\begin{equation}
\label{eq:J-with-3wd}
J = 2\omega_d r_1^2\sin k_1 + 6\omega_d r_3^2\sin 3k_1.
\end{equation}
We evaluated this numerically and in Fig.~\ref{fig:k1-vs-wd} plot this as a function of $\omega_d$ for fixed parameters $F_d=10.0$ and $N=100$. We have also shown comparisons with the results obtained from direct simulations and from the first harmonic result in Eq.~\eqref{eq:j_form} and see that at higher frequencies the third harmonic computation gives a better agreement to the simulations. The inset in Fig.~\ref{fig:k1-vs-wd} shows the dependence of the wavenumber $k_1$ on $\omega_d$ and in this case we see no noticeable difference between the first and third harmonic computations.   

\begin{figure}[t]
\includegraphics[width=\linewidth]{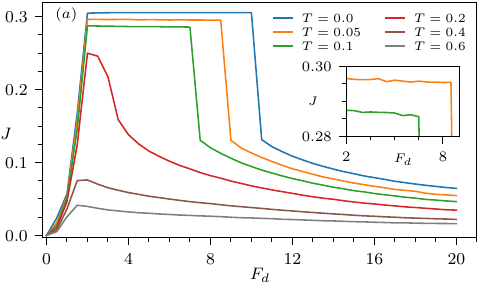}
\includegraphics[width=\linewidth]{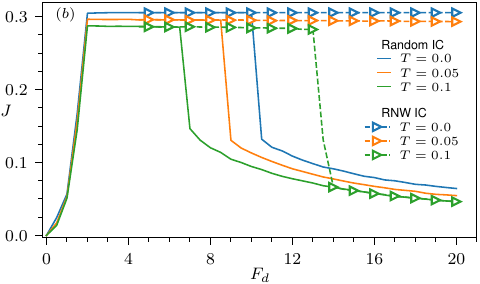}
\caption{(color online) {\bf PBS setup ($T_L=T_R=T$):}
Variation of steady state current $J$ with driving amplitude $F_d$ at different bath temperatures for system size $N=500$ and $\omega_d=1.5$.  (a) In this case, the system was initialized with Random ICs. We observe that for small enough temperatures the sharp transitions observed for $T=0$ still persist, but over a smaller range and a reduced value of $J$ in the RNW regime. However, in this regime, the current $J$ has a small slope (see inset) in contrast to the $T=0$ case. With increasing $T$, the RNW regime ceases to exist. 
(b) Here we explore if the flat regime corresponding to RNW mode persists beyond the second transition point $F_2$ even at non-zero temperature as happens for $T=0$ (see Fig.~\ref{fig:f1f2-rnw-crit-perturb}). For this, the system is again initialized with RNW IC and we observe that the RNW phase continues to exist beyond $F_2$ (triangles). However, the range becomes smaller as $T$ increases. The solid lines corresponds to the profiles obtained from random IC [as plotted in (a)].
}
\label{fig:rnw-at-finite-temp}
\end{figure}

\section{\label{sec:T!=0} Effects of finite temperature boundary baths}

So far, we have studied the system in the absence of any thermal noise. In this section, we study the effect of thermal noise on the observed transitions and the RNW mode. Apart from the PBS setup, we also consider the KLO setup, where the boundary damping and the form of driving are somewhat different. Thus, we consider  the set of equations given in Eq.~\eqref{eq:eom_s} with $\lambda=1$, for three cases: (A) PBS case ---  $\gamma_L=\gamma_R=1$, $T_L=T_R=T\neq 0$; (B) KLO case --- $\gamma_L=0$, $\gamma_R=1$ and $T_R=T\neq 0$, $F_d=A/N^{1/2}$; (C)  $\gamma_L=0$, $\gamma_R=1$ and $T_R=T\neq 0$ but with system-size independent $F_d$.

\subsection{PBS setup at non-zero bath temperatures}
\label{subsec:pbs}
In this case, the equation of motion in Eq.~\eqref{eq:eom_T0} are modified to,
\begin{align}
\label{eq:eom_T}
\ddot{q_\ell} 
&= -q_\ell - q_\ell^3 + (q_{\ell+1}+q_{\ell-1}-2 q_\ell) \nonumber\\
&\qquad\qquad\quad
+\delta_{\ell, 1}(-\dot{q_\ell} +\sqrt{2 T} \eta_L
+ F_d\cos\omega_d t)\nonumber\\
&\qquad\qquad\quad
+\delta_{\ell, N}(-\dot{q_\ell} +\sqrt{2T} \eta_R(t)),
\end{align}
for $\ell =1,2,\ldots,N$ with $q_0=q_{N+1}=0$. In  Fig.~\ref{fig:rnw-at-finite-temp}$a$, we present simulation results for the steady state current as a function of the driving force for fixed driving frequency $\omega_d=1.5$ and for a set of temperatures.  
Surprisingly, we find that at the lowest observed temperature ($T=0.1$), the form of the current dependence on force is similar to the zero-noise case --- we still see sharp transitions, though  the plateau region now has a decreased range. The current has a smaller value and  in fact we observe a small slope (see inset of Fig.~\ref{fig:rnw-at-finite-temp}$a$). At higher temperatures, $T>0.2$ we do not see the plateau region, implying that the  RNW mode is  either not present or has a negligible effect.   
\begin{figure}[h]
\centering
\includegraphics[width=\linewidth]{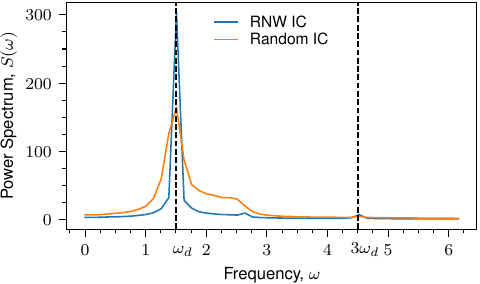}
\caption{(color online) {\bf PBS setup ($T_L=T_R=T$):} Average power spectrum of the positions of the particles in a chain with $N=500$,  $\omega_d=1.5$ and $F_d=10.0$ at bath temperature $T=0.01$. The power spectrum, $S(\omega)= \sqrt{\langle|q_\ell(\omega)|^2\rangle}$, is computed for each particle and then an average $\langle ... \rangle$ is taken over all the particles. The blue line corresponds to the RNW IC and the orange line corresponds to the random IC. For both cases, $S(w)$ has a large peak at $w=w_d$ and a very tiny secondary peak at $w=3w_d$ (indicated by the black dashed vertical lines). The peak for the RNW case is very sharp, whereas the peak is broadened for the random IC case.}
\label{fig:avg-freq-resp}
\end{figure}

\begin{figure}[h]
\includegraphics[width=\linewidth]{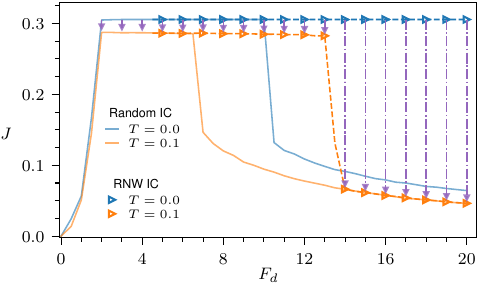}
\caption{ {\bf PBS setup ($T_L=T_R=T$):} Here we test the stability of the zero temperature RNW mode beyond $F_2$ by observing how the current in the steady state get affected upon addition of thermal noise with $T=0.1$. 
We notice that till some critical $F_d$, which is the same as in Fig.~\ref{fig:rnw-at-finite-temp}b at $T=0.1$, the system remains in the RNW phase characterized by a lower value of current. For larger $F_d$, the addition of noise causes a transition to the chaotic phase (indicated by vertical arrows). This is further indication of the stability of the RNW phase even at small finite temperatures. In this plot we take $N=500$ and $\omega_d=1.5$.
}
\label{fig:unique-ss}
\end{figure}

We next explore  whether (for temperatures $T\leq 0.1$) multiple nonequilibrium steady states exist even in the presence of noise, beyond the transition point $F_2$. For this we again start from initial conditions taken when the system is in the plateau region and then we increase the force in small steps. Again, somewhat surprisingly, we find that the plateau region has an extended domain of stability, as shown in Fig.~\ref{fig:rnw-at-finite-temp}$b$. 
However, now the extended region of stability shows a clear decrease with  increasing temperature. All this implies that there are two NESS states, one  corresponding to the RNW mode, with a sharp power spectrum peaked at $\omega_d$ (see Fig.~\ref{fig:avg-freq-resp}),  and the other to the low current chaotic state, with a broad power spectrum. As further evidence of the extended stability and existence of a second NESS, we show in Fig.~\ref{fig:unique-ss} the results of simulations where  the noise is switched on after starting from the zero noise RNW initial condition. We see that over a range of $F_d$ (which is the same as seen in Fig.~\ref{fig:rnw-at-finite-temp}$b$), the system transits to the RNW state.  It is possible that the RNW state is a metastable state, but our simulations do not see a transition to the low current state even at very long times. 

It was observed by PBS~\cite{prem2023} that in the chaotic phase for $F_d >F_2$, while the current obeyed the Fourier behavior $J \sim 1/N$, the temperature profile in the bulk of the chain was far from that expected from Fourier's law.  

\begin{figure}[h]
\includegraphics{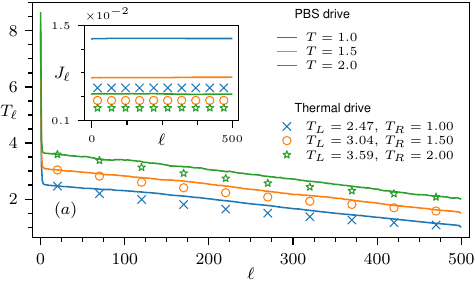}
\includegraphics{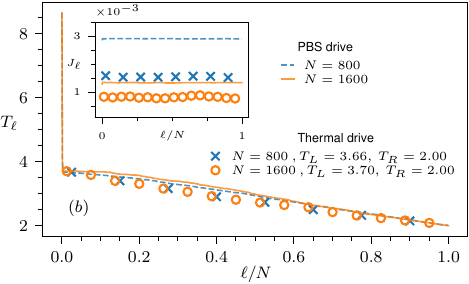}
\caption{{\bf PBS setup ($T_L=T_R=T$):}  Comparison of temperature and current (inset) profiles obtained from periodic driving (solid lines) and equivalent thermal drive (points).
In (a) we show these profiles for  $N=500$ at three different temperatures $T=1,~1.5$ and $2$. We see that while the temperature profiles show reasonable agreement,  the agreement for the current profiles  is not so good but gets better with increasing temperature. In (b) we plot the profiles at $T=2$ for two  system sizes $N=800$ and $N=1600$ and we observe that the temperature profiles have converged and the agreement of the current profiles with the equivalent thermal drive case improves with increasing $N$.
}
\label{fig:PBS-vs-thermal}
\end{figure}

We now ask  if things are different in the presence of boundary noise (at high temperatures) and in particular if the bulk temperature profiles follow Fourier's law. To check this, we first evaluate the temperature profile in the driven KG chain and the results are shown in  Fig.~\ref{fig:PBS-vs-thermal}$a$ at different bath temperatures for $N=500$ and in Fig.~\ref{fig:PBS-vs-thermal}$b$ for different $N$ at fixed temperature $T=2.0$. The profiles show a jump at the driven end and then a slowly varying profile. We compare these profiles with those obtained from simulations where there is no external force but only an imposed temperature gradient (see end of Sec.~\ref{sec:model}). Specifically, we consider a chain of length $N-i_s$ and fix the temperature at the first site to be that of  the temperature at the $i_s$ site of the driven chain (we choose $i_s=20$, i.e, far from the left boundary) and the temperature of the right end to be the same as in the driven chain. In Fig.~\ref{fig:PBS-vs-thermal}, we compare the temperature profile and the local current profile (inset)  of the thermally driven chain  with the corresponding segment in the periodically driven chain. We find that there is reasonable agreement which improves with increasing $N$, implying that the transport in the bulk is in accordance to the Fourier diffusion equation. 

\subsection{KLO setup at non-zero bath temperatures}
We now discuss the KLO setup with dissipation only at one end, i.e, $\gamma_L=0, \gamma_R =1$ and   a forcing with amplitude $F_d=A/\sqrt{N}$ where $A$ is a constant. In addition, we have thermal noise at the right end.  One of the main results of \cite{komorowski2023}, for the case where the system is a  harmonic chain with an energy conserving  stochastic dynamics, was to show that transport in the chain is diffusive. Here we explore if a similar  transport behavior is observed in the KG chain with Hamiltonian dynamics and a similar driving protocol.
\begin{figure}[h]
\includegraphics{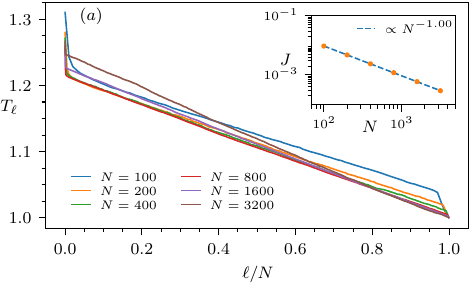}
\includegraphics{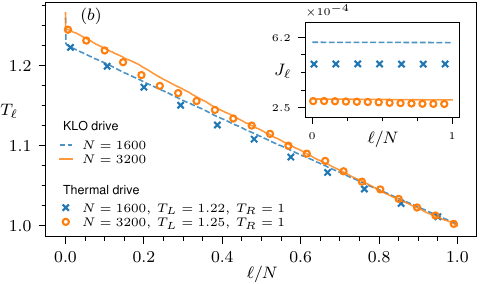}
\caption{ {\bf KLO setup ($\gamma_L=0,~T_R=1.0$)}: (a) In this case, we plot the temperature profiles for different system sizes with driving amplitude $F_d=A/\sqrt{N}$ with $A=1$ and driving frequency $\omega_d=1.5$. We observe that the temperature profiles seems to converge to a limiting form for large $N$. The inset verifies the Fourier scaling of the current with system size. (b) Comparison of temperature profile and current profile (inset) between periodic driving case  with $A=1$ and the equivalent thermally driven chain for $N=1600$ and $N=3200$. We again observe good agreement between the temperature profiles for each, $N$ while the agreement for the current profiles gets improved with increasing $N$.}
\label{fig:Tx-KLO-scaled-A1}
\end{figure}
In Fig.~\ref{fig:Tx-KLO-scaled-A1}$a$, we present  results  for the temperature profile and (in the inset) the size-dependence of the current. For the driving force of the form $F_d=A/\sqrt{N}$, we took  $A=1$,  $\omega_d=1.5$ and  $T=1.0$.  We  see that the current scaling, $J \sim 1/N$, is  consistent with diffusive transport (see inset of Fig.~\ref{fig:Tx-KLO-scaled-A1}$a$). With increasing $N$, the  temperature profiles show a slow convergence to a limiting profile, as one might expect from the results of  Ref.~\cite{komorowski2023}.  To check the validity of the diffusion equation,  we compare the temperature and current profiles (in the bulk)  of the periodically driven chain with that of a thermally driven chain with temperatures at the boundaries made identical at the  end points (see Sec.~\eqref{subsec:pbs} for details). The results are shown in Fig.~\ref{fig:Tx-KLO-scaled-A1}$b$. We see that the agreement between the profiles of the thermal and periodically driven cases gets better with increasing system size, indicating again that Fourier's law is satisfied in the bulk of the system.

\subsection{Case with zero dissipation at driving end and unscaled force}
\begin{figure}[h]
\includegraphics{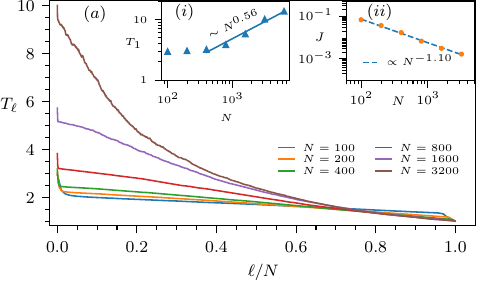}
\includegraphics{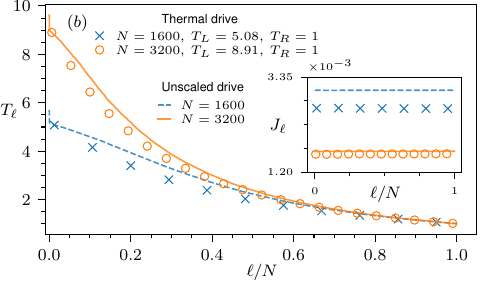}
\caption{{\bf Unscaled drive ($\gamma_L=0,~T_R=1.0$ and $F_d=1.0)$:} (a) Here we show the temperature profiles  at different system sizes with  $\omega_d=1.5$ and fixed driving amplitude $F_d=1.0$. In this case the temperature profiles do not converge with increasing $N$ and the temperatures of the particles at the left end keep increasing with $N$. In the inset (i), we show that the kinetic temperature of the first particle 
 grows as $N^{0.56}$. 
 In the inset (ii), we observe that the current seems to show Fourier scaling with $N$. (b) 
Comparison of the temperature profiles and current profiles (inset) in the periodically driven case  and the equivalent thermally driven chain for $N=1600$ and $N=3200$. We again observe good agreement between the temperature profiles for each $N$ while the agreement for the current profiles gets improved with increasing $N$.
}
\label{fig:Tx-KLO-unscaled-A1}
\end{figure}

Finally, we consider again setup with dissipation and thermal noise only at the right end and  with no system-size scaling of the driving force, i.e, with a constant drive $F_d=1$. The temperature profile for different system-size  are shown in  Fig.~\ref{fig:Tx-KLO-unscaled-A1}$a$ and in the inset we again observe the Fourier scaling for the current, i.e,  $J\propto 1/N$. In this case, the temperature profile does not seem to converge with increasing $N$ and in fact the kinetic temperature of the first particle grows approximately as $N^{1/2}$ (see inset of \fig{Tx-KLO-unscaled-A1}$a$). In Fig.~\ref{fig:Tx-KLO-unscaled-A1}$b$, we once again compare the temperature and current profiles of the periodically driven chain with the purely thermally driven chain for which the temperature at the left end is fixed following the same procedure as discussed at the end of Sec.~\ref{subsec:pbs}. The good agreement of both the temperature and current profiles between the two methods at the largest system size, suggests that Fourier's law is satisfied in the bulk of the chain in this case also.


\section{\label{sec:conclusion} Conclusion}
We revisited the recently studied problem of transport through KG chain \cite{prem2023} that is periodically driven at one of the boundaries and with dissipation at both ends and presented an improved understanding of the Resonant Nonlinear Wave (RNW) mode. We establish that the RNW mode has an extended domain of stability in the driving parameters ($F_d,\omega_d)$ space and that, in certain parameter regions, there can be multiple attractors. At lowest order, the RNW mode is a periodic wave at frequency $w_d$. We provided a quantitative estimate of the corrections coming from the third harmonic contributions.  It is interesting to note that two stable attractors (bistability)  were also observed in  Ref.~\cite{khomeriki2004} for the case of a periodically driven Fermi-Pata-Ulam chain.  
Some notable differences with that study are the facts that transmission was observed only for driving frequency outside the phonon bandwidth, and transport was via moving solitons.  

Finally, we looked at the effect of thermal noise on the RNW mode and also on transport properties. We found that at low temperatures, the features of RNW mode survives, while at high temperature the transport is in accordance with diffusion equation and Fourier's law with some effective temperature at the (periodically) driven end. We point out the effect of boundary conditions on the effective (purely) thermally driven chain problem by studying two setups, including the one recently studied in \cite{pedro2023klo}.

The robustness of the RNW mode means that it is amenable to being observed in experimental setups such as those in ~\cite{fitzpatrick2017,fedorov2021} or in macroscopic mass-spring chains such as the one studied in ~\cite{watanabe2018}. 

There remain several  open questions. For the noiseless case, establishing the existence of multiple stable attractors, determining the transition points $F_1,F_2$, and  the analytic determination of the wave number $k$ of the RNW mode are interesting problems. In the presence of thermal noise, the naive expectation would be that the system goes to a unique time periodic Floquet NESS~\cite{bukov2015,higashikawa2018floquet}, while our results indicate the existence of multiple steady states --- how does one understand this ?
Finally, for the case where the noise strength is large, an interesting problem is to establish that  transport is diffusive, that Fourier's law is satisfied in the bulk of the chain and that the periodic driving can be replaced by an effective boundary condition.  

\begin{acknowledgments}
We thank Vir Bulchandani, Shiva Darshan, Sergej Flach, Joel Lebowitz, Stefano Olla and  Abhinav Prem for useful discussions.  AK would like to acknowledge the support of DST, Government of India Grant under Project No. ECR/2017/000634 and the MATRICS grant MTR/2021/000350 from the SERB, DST, Government of India. We acknowledge the Department of Atomic Energy, Government of India, for their support under Project No. RTI4001.  We acknowledge  the ICTS program "Periodically and quasi-periodically driven complex systems"  (code: ICTS/pdcs2023/6) for enabling very useful discussions. 
\end{acknowledgments}
\newpage
\appendix

\bibliography{refs2023}

\end{document}